\documentclass[11pt]{article}

\usepackage{indentfirst}
\usepackage{graphicx}
\usepackage{longtable}
\usepackage{supertabular}
\usepackage{amsmath}
\usepackage{amssymb}
\usepackage{color}
\usepackage{amstext}
\usepackage{cases}
\usepackage{float}
\usepackage{threeparttable}
\usepackage{booktabs}
\usepackage[margin=.8in]{geometry}
\usepackage{setspace}
\usepackage{multirow}
\usepackage{array}
\usepackage{mathrsfs}

\begin{document}

\title{{\Large\bf {Reformulating the Value Restriction and the Not-Strict Value Restriction in Terms of Possibility Preference Map}}}

\author{Fujun Hou \thanks{Email: houfj@bit.edu.cn.} \\
School of Management and Economics\\
Beijing Institute of Technology\\
Beijing, China, 100081
}
\date{\today}
\maketitle

\begin{abstract}
In social choice theory, Sen's value restriction and Pattanaik's not-strict value restriction are both attractive conditions for testing social preference transitivity and/or non-empty social choice set existence. This article introduces a novel mathematical representation tool, called possibility preference map (PPM), for weak orderings, and then reformulates the value restriction and the not-strict value restriction in terms of PPM. The reformulations all appear elegant since they take the form of minmax.

\end{abstract}

\setlength{\unitlength}{1mm}

\section{Introduction}

Ever since Arrow's seminal work of Impossibility Theorem (Arrow, 1950, 1951), seeking conditions for transitive social preference and non-empty social choice set under the majority decision rule has been a fundamental topic in social choice theory. So far, a number of restrictions have been identified (Arrow,1951; Black,1958; Inada,1964,1969; Ward, 1965; Sen, 1966; Sen \& Pattanaik, 1969; Pattanaik, 1970; Duggan,2016; and many others). Among those already identified, Sen's value restriction (Sen, 1966) and Pattanaik's not-strict value restriction (Pattanaik, 1970) are attractive since they both cover some conditions precede them. The objective of this paper is to reformulate these two value-relevant restrictions in another way.

The individuals preferences are assumed as weak orderings over a finite alternative set. A novel mathematical tool, called possibility preference map (PPM), is introduced for representing the weak orderings. A PPM is a square matrix, whose element indicates the possibility of a certain alternative being ranked at a certain position. We then reformulate the value restriction (VR) and the not-strict value restriction (NSVR) in terms of the PPM.

\section{Preliminary}

\subsection{Social transitivity and social choice set}

Each individual is assumed to hold a weak ordering as her preference over a finite set of alternatives. Following Sen (1966) and Pattanaik (1970), individual $i$'s preferences over two alternatives $x$ and $y$ such as "$x$ is at least as good as $y$", "$x$ is preferred to $y$" and "$x$ is indifferent to $y$" are denoted by $x R_i y$, $x P_i y$ and $x I_i y$, respectively. Let $R$, $P$ and $I$ denote the corresponding social preferences. Let $N(xR_iy)$, $N(xP_iy)$ and $N(xI_iy)$ denote the numbers of individuals of those holding $x R_i y$, $x P_i y$ and $x I_i y$, respectively. The majority decision rule means:
\begin{itemize}
\item [$\diamond$] $xRy$ if and only if $N(xRy)\geq N(yRx)$,
\item [$\diamond$] $xPy$ if and only if $N(xRy)> N(yRx)$,
\item [$\diamond$] $xIy$ if and only if $N(xRy)= N(yRx)$.
\end{itemize}

If "($xRy$ \& $yRz$) $\rightarrow$ $xRz$" holds for every triple of alternatives, then the social transitivity is guaranteed (Sen, 1966). A social choice set of alternative set $S$ is defined by a subset $C(S)$ of $S$ such that every element in this subset is socially at least as good as every element in $S$ (Sen \& Pattanaik,1969).

\subsection{Value restriction (VR) and not-strict value restriction (NSVR)}

The value restriction (VR) is established over alternative triples (Sen, 1966). Regarding an alternative in a triple, an individual associates, according to her preference over the triple, a value which can be "best", "worst" or "medium" with the alternative in consideration. To develop the VR, Sen distinguishes concerned individuals between unconcerned individuals. If an individual is indifferent between all the alternatives, then she is called a unconcerned individual. Otherwise, she is a concerned one (Sen, 1966).

\textbf{Value restriction (VR)} A set of individual preferences over a triple is called value restricted when there exist one alternative and one value such that none of the individuals think that the alternative takes that value.

Sen (1966) proved a theorem of transitive social preference, that the majority decision rule yields a transitive social ordering when the preferences of concerned individuals over every triple of alternatives satisfy the VR and the number of concerned individuals over every triple is odd.

Sen's value-restricted preference pattern combines the single-peaked preference pattern of Arrow (1951) and Black (1958), the single-caved preference pattern of Inada (1964), and the two-group-separated preference pattern of Inada (1964).

The not-strict value restriction (NSVR) is of conditions for non-empty social choice set. It involves three patterns respectively called not-strict best (NSB), not-strict worst (NSW) and not-strict medium (NSM). They are defined quite similar to that of VR but with the strict preferences excluded either. Pattanaik (1970) proved two theorems with one confirming that the choice set is nonempty provided that the NSB value restriction is fulfilled by each of the alternative triples, or that the NSW value restriction is fulfilled by each of the alternative triples; and the other confirming that the choice set of a triple is nonempty provided that the NSM value restriction is fulfilled by the triple. Pattanaik (1970) also pointed out that NSM value restriction is only sufficient for alternative set which includes not more than 3 alternatives.
Pattanaik's NSVR covers the case considered by Dummet and Farquharson (1961).

\section{Possibility preference map (PPM)}

The preference map (PM) is a mathematical tool for representing weak orderings over a finite alternative set. It is established on two assumptions: (a) each alternative corresponds to a set which contains the alternative's ranking position or positions; and (b) The alternatives in a tie are tied together occupying consecutive positions (Hou, 2015a, 2015b).

Suppose that the alternative set is $S=\{x_1,x_2,\ldots,x_m\}$ where $1<m<+\infty$. A PM is defined as follows (Hou, 2015a, 2015b; Hou \& Triantaphyllou, 2019).

\textbf{Preference map (PM)} The PM corresponding to individual $j$'s weak ordering over the alternative set $S=\{x_1,x_2,\ldots,x_m\}$ is defined by a preference sequence $PM^{(j)}=[PM_i^{(j)}]_{m\times 1}$ such that
$$PM_i^{(j)}=\{|A_i^{(j)}|+1,|A_i^{(j)}|+2,\ldots,|A_i^{(j)}|+|B_i^{(j)}|\},\eqno(1)$$
where the notation $|\centerdot|$ stands for the cardinality of a set;
$A_i^{(j)}$ stands for the \textit{predominance set} of alternative $x_i$ according to individual $j$'s preference, i.e., $A_i^{(j)}=\{x_q\mid x_q\in S, x_q P_j x_i\}$; and $B_i^{(j)}$ stands for the \textit{indifference set} of alternative $x_i$, i.e., $B_i^{(j)}=\{x_q\mid x_q\in S,x_q I_j x_i\}$.

For example, the PMs corresponding to the orderings of $x_1Px_2Px_3$, $x_1Ix_2Px_3$ and $x_1Ix_2Ix_3$ are
$$
\begin{array}{c}
{\begin{array}{cc}
{\begin{array}{c} x_1\\
x_2\\
x_3
\end{array}}&
{ \left[\begin{array}{c}
\{1\}\\
\{2\}\\
\{3\}\end{array}\right],
\left[\begin{array}{c}\{1,2\}\\
\{1,2\}\\
\{3\}\end{array}\right]\mbox{~and~}
\left[\begin{array}{c}\{1,2,3\}\\
\{1,2,3\}\\
\{1,2,3\}\end{array}\right],
}
\end{array}}
\end{array}
$$
respectively.

To characterize not-strict preference more intuitively, we define a novel representation tool for weak orderings based on the PM.

\textbf{Possibility preference map (PPM)} Let $PM=[PM_i]_{m\times 1}$ be a PM corresponding to a weak ordering. The PPM corresponding to the weak ordering is defined by a $m\times m$ matrix $[PPM_{i,k}]_{m\times m}$ such that
$$
{PPM_{i,k}}=\begin{cases}
\frac{1}{|PM_i|}, & \text{if $k\in PM_i$},\\
0, & \text{otherwise}.
\end{cases}
\eqno(2)$$

One can see from Eq.(2) that the PPM implies a specific exposition of the assumption for a PM that the alternatives in a tie will occupy the consecutive positions with equal possibility, and the possibility level is $\frac{1}{|PM_i|}$. This sort of strategy can be traced back at least as early as Black (1976) when to assign Borda scores to those alternatives in a tie.

To illustrate the concept of PPM, we reconsider the above three orderings. Their corresponding PPMs are
$$
\begin{array}{c}
\begin{array}{c}{\hspace{0.5cm}\begin{array}{cccc}B\hspace{0.3cm}M\hspace{0.3cm}W
&\hspace{0.6cm}B\hspace{0.4cm}M\hspace{0.4cm}W &\hspace{0.9cm}B\hspace{0.5cm}M\hspace{0.5cm}W
\end{array}}\\
{\begin{array}{cc}
{\begin{array}{c} x_1\\
x_2\\
x_3
\end{array}}&
{ \left[\begin{array}{ccc}
1 &  0 &  0 \\
0 &  1 &  0 \\
0 &  0 &  1
\end{array}\right],
\left[\begin{array}{ccc}
1/2 &  1/2 &  0 \\
1/2 &  1/2 &  0 \\
0 &  0 &  1
\end{array}\right],
\left[\begin{array}{ccc}
1/3 &  1/3 &  1/3 \\
1/3 &  1/3 &  1/3 \\
1/3 &  1/3 &  1/3
\end{array}\right],
}
\end{array}}
\end{array}
\end{array}
$$
respectively.

Many interesting properties of the PPM might be concluded. For an individual's PPM with respect to a triple $(x_1,x_2,x_3)$, however, the following three desirable characteristics are sufficient for our discussion.
\begin{itemize}
\item if $PPM_{i,k}=1$ then the individual would like to assign a strict value (best, worst or medium, differentiated by $k$) to alternative $x_i$.
\item if $0<PPM_{i,k}<1$ then the individual would like to assign a not-strict value (best, worst or medium, differentiated by $k$) to alternative $x_i$.
\item if $PPM_{i,k}=0$ then the individual does not like to assign position $k$ to alternative $x_i$.
\end{itemize}

\section{Reformulation of value-relevant restriction}

\subsection{Reformulation of VR}

From Section 3, we know that, if the $(i,k)$ element of an individual's PPM takes the value 0, then alternative $x_i$ will not be ranked at position $k$. Thanks to this sort of desirable information possibly conveyed by the PPM, we are able to reformulate the value restriction (VR).

Suppose the individual set is $N=\{1,2,\ldots,n\}$ with $1<n<+\infty$, and the individuals' PPMs over a triple $(x_1,x_2,x_3)$ are $PPM^{(j)}=[PPM_{i,k}^{(j)}]_{3\times 3}$, $j=1,2,\ldots,n$.

\textbf{VR in terms of PPM} A set of $n$ individual preferences over triple $(x_1,x_2,x_3)$ is value restricted if and only if the following condition is satisfied
$$\min_{i\in \{1,2,3\}}\min_{k\in \{1,2,3\}}\max_{j\in N}\left\{PPM_{i,k}^{(j)}\right\}=0.\eqno(3)$$
Particularly,
\begin{itemize}
\item [$\lozenge$] \textbf{Not Best (NB) in terms of PPM} A set of $n$ individual preferences over triple $(x_1,x_2,x_3)$ is best value restricted if and only if the following condition is satisfied
$$\min_{i\in \{1,2,3\}}\max_{j\in N}\left\{PPM_{i,k}^{(j)}\right\}=0, k=1.\eqno(4)$$
\item [$\lozenge$] \textbf{Not Medium (NM) in terms of PPM} A set of $n$ individual preferences over triple $(x_1,x_2,x_3)$ is medium value restricted if and only if the following condition is satisfied
$$\min_{i\in \{1,2,3\}}\max_{j\in N}\left\{PPM_{i,k}^{(j)}\right\}=0, k=2.\eqno(5)$$
\item [$\lozenge$] \textbf{Not Worst (NW) in terms of PPM} A set of $n$ individual preferences over triple $(x_1,x_2,x_3)$ is worst value restricted if and only if the following condition is satisfied
$$\min_{i\in \{1,2,3\}}\max_{j\in N}\left\{PPM_{i,k}^{(j)}\right\}=0, k=3.\eqno(6)$$
\end{itemize}

\subsection{Reformulation of NSVR in terms of PPM}

Similar to subsection 4.1 but making use of the first two desirable characteristics listed in the end of Section 3, we reformulate the not-strict value restriction (NSVR) as follows.

\begin{itemize}
\item [$\lozenge$] \textbf{Not-strict Best (NSB) in terms of PPM} A set of $n$ individual preferences over triple $(x_1,x_2,x_3)$ is not-strict best value restricted if and only if the following condition is satisfied
$$\min_{i\in \{1,2,3\}}\max_{j\in N}\left\{PPM_{i,k}^{(j)}\right\}<1, k=1.\eqno(7)$$
\item [$\lozenge$] \textbf{Not-strict Medium (NSM) in terms of PPM} A set of $n$ individual preferences over triple $(x_1,x_2,x_3)$ is not-strict medium value restricted if and only if the following condition is satisfied
$$\min_{i\in \{1,2,3\}}\max_{j\in N}\left\{PPM_{i,k}^{(j)}\right\}<1, k=2.\eqno(8)$$
\item [$\lozenge$] \textbf{Not-strict Worst (NSW) in terms of PPM} A set of $n$ individual preferences over triple $(x_1,x_2,x_3)$ is not-strict worst value restricted if and only if the following condition is satisfied
$$\min_{i\in \{1,2,3\}}\max_{j\in N}\left\{PPM_{i,k}^{(j)}\right\}<1, k=3.\eqno(9)$$
\end{itemize}

\section{Illustration}

For illustrative purpose, we test the reformulated restrictions on an example which was considered by Sen (1966).

\textbf{Example 1} Suppose the alternative set is $S=\{w,x,y,z\}$, and the individual preference orderings are
$$\begin{array}{ll}
 \text{Individual 1}: & w I_1 x P_1 y P_1 z,\\
 \text{Individual 2}: & x I_2 w P_2 z P_2 y,\\
 \text{Individual 3}: & z I_3 x P_3 y P_3 w,\\
 \text{Individual 4}: & z P_4 y I_4 x P_4 w,\\
 \text{Individual 5}: & z P_5 y P_5 x P_5 w.
\end{array}
$$
As pointed out by Sen (1966), there are 4 possible alternative triples and they are all value restricted, and thus the majority decision yields a transitive social preference ordering, that is, $xIzPyPw$. Moreover, the social choice set is nonempty, and it is $\{x,z\}$. In this example the individual set is $N=\{1,2,3,4,5\}$.

Here we test the preferences' value-relevant restriction characteristic.

(1) With regard to triple $(w,x,y)$, the PMs corresponding to the five individuals are
$$
\begin{array}{c}
{\begin{array}{cc}
{\begin{array}{c} w\\
x\\
y
\end{array}}&
{ \left[\begin{array}{c}
\{1,2\}\\
\{1,2\}\\
\{3\}\end{array}\right],
\left[\begin{array}{c}\{1,2\}\\
\{1,2\}\\
\{3\}\end{array}\right],
\left[\begin{array}{c}\{3\}\\
\{1\}\\
\{2\}\end{array}\right],
\left[\begin{array}{c}\{3\}\\
\{1,2\}\\
\{1,2\}\end{array}\right],
\left[\begin{array}{c}\{3\}\\
\{2\}\\
\{1\}\end{array}\right].
}
\end{array}}
\end{array}
$$
Accordingly, their respective PPMs are
$$
\begin{array}{c}
\begin{array}{c}
{\begin{array}{cc}
{\begin{array}{c} w\\
x\\
y
\end{array}}&
{ \left[\begin{array}{ccc}
1/2 &  1/2 &  0 \\
1/2 &  1/2 &  0 \\
0 &  0 &  1
\end{array}\right],
\left[\begin{array}{ccc}
1/2 &  1/2 &  0 \\
1/2 &  1/2 &  0 \\
0 &  0 &  1
\end{array}\right],
\left[\begin{array}{ccc}
0 &  0 &  1 \\
1 &  0 &  0 \\
0 &  1 &  0
\end{array}\right],
\left[\begin{array}{ccc}
0 &  0 &  1 \\
1/2 &  1/2 &  0 \\
1/2 &  1/2 &  0
\end{array}\right],
\left[\begin{array}{ccc}
0 &  0 &  1 \\
0 &  1 &  0 \\
1 &  0 &  0
\end{array}\right].
}
\end{array}}
\end{array}
\end{array}
$$
\begin{itemize}
\item
The set of preferences over triple $(w,x,y)$ is worst value restricted (i.e., single-peaked) due to $\max_{j\in N}\left\{PPM_{2,3}^{(j)}\right\}=0$. Specifically, $x$ is NW restricted.

\item The set of preferences over triple $(w,x,y)$ is NSW value restricted due to $\max_{j\in N}\left\{PPM_{2,3}^{(j)}\right\}=0$, NSB value restricted due to $\max_{j\in N}\left\{PPM_{1,1}^{(j)}\right\}=1/2$ and NSM value restricted due to\\ $\max_{j\in N}\left\{PPM_{1,2}^{(j)}\right\}=1/2$. Specifically, $x$ is NSW restricted, while $w$ is both NSB restricted and NSM restricted.
\end{itemize}

(2) With regard to triple $(w,y,z)$, the PMs corresponding to the five individuals are
$$
\begin{array}{c}
{\begin{array}{cc}
{\begin{array}{c} w\\
y\\
z
\end{array}}&
{ \left[\begin{array}{c}
\{1\}\\
\{2\}\\
\{3\}\end{array}\right],
\left[\begin{array}{c}\{1\}\\
\{3\}\\
\{2\}\end{array}\right],
\left[\begin{array}{c}\{3\}\\
\{2\}\\
\{1\}\end{array}\right],
\left[\begin{array}{c}\{3\}\\
\{2\}\\
\{1\}\end{array}\right],
\left[\begin{array}{c}\{3\}\\
\{2\}\\
\{1\}\end{array}\right].
}
\end{array}}
\end{array}
$$
Accordingly, their respective PPMs are
$$
\begin{array}{c}
\begin{array}{c}
{\begin{array}{cc}
{\begin{array}{c} w\\
y\\
z
\end{array}}&
{ \left[\begin{array}{ccc}
1 &  0 &  0 \\
0 &  1 &  0 \\
0 &  0 &  1
\end{array}\right],
\left[\begin{array}{ccc}
1 &  0 &  0 \\
0 &  0 &  1 \\
0 &  1 &  0
\end{array}\right],
\left[\begin{array}{ccc}
0 &  0 &  1 \\
0 &  1 &  0 \\
1 &  0 &  0
\end{array}\right],
\left[\begin{array}{ccc}
0 &  0 &  1 \\
0 &  1 &  0 \\
1 &  0 &  0
\end{array}\right],
\left[\begin{array}{ccc}
0 &  0 &  1 \\
0 &  1 &  0 \\
1 &  0 &  0
\end{array}\right].
}
\end{array}}
\end{array}
\end{array}
$$
\begin{itemize}
\item
The set of preferences over triple $(w,y,z)$ is best value restricted (i.e., single-caved) due to\\ $\max_{j\in N}\left\{PPM_{2,1}^{(j)}\right\}=0$ and medium value restricted (i.e., two-group-separated) due to\\ $\max_{j\in N}\left\{PPM_{1,2}^{(j)}\right\}=0$. Specifically, $w$ is NM restricted and $y$ is NB restricted.

\item Because the individuals' preferences over $(w,y,z)$ are all strict preferences, thus, value restriction implies not-strict value restriction (the reverse does not necessarily hold). Hence we know that $w$ is NSM restricted and $y$ is NSB restricted.
\end{itemize}

(3) With regard to triple $(w,x,z)$, the PMs corresponding to the five individuals are
$$
\begin{array}{c}
{\begin{array}{cc}
{\begin{array}{c} w\\
x\\
z
\end{array}}&
{ \left[\begin{array}{c}
\{1,2\}\\
\{1,2\}\\
\{3\}\end{array}\right],
\left[\begin{array}{c}\{1,2\}\\
\{1,2\}\\
\{3\}\end{array}\right],
\left[\begin{array}{c}\{3\}\\
\{1,2\}\\
\{1,2\}\end{array}\right],
\left[\begin{array}{c}\{3\}\\
\{2\}\\
\{1\}\end{array}\right],
\left[\begin{array}{c}\{3\}\\
\{2\}\\
\{1\}\end{array}\right].
}
\end{array}}
\end{array}
$$
Accordingly, their respective PPMs are
$$
\begin{array}{c}
\begin{array}{c}
{\begin{array}{cc}
{\begin{array}{c} w\\
x\\
z
\end{array}}&
{ \left[\begin{array}{ccc}
1/2 &  1/2 &  0 \\
1/2 &  1/2 &  0 \\
0 &  0 &  1
\end{array}\right],
\left[\begin{array}{ccc}
1/2 &  1/2 &  0 \\
1/2 &  1/2 &  0 \\
0 &  0 &  1
\end{array}\right],
\left[\begin{array}{ccc}
0 &  0 &  1 \\
1/2 &  1/2 &  0 \\
1/2 &  1/2 &  0
\end{array}\right],
\left[\begin{array}{ccc}
0 &  0 &  1 \\
0 &  1 &  0 \\
1 &  0 &  0
\end{array}\right],
\left[\begin{array}{ccc}
0 &  0 &  1 \\
0 &  1 &  0 \\
1 &  0 &  0
\end{array}\right].
}
\end{array}}
\end{array}
\end{array}
$$
\begin{itemize}
\item
The set of preferences over triple $(w,x,z)$ is worst value restricted (i.e., single-peaked) due to\\ $\max_{j\in N}\left\{PPM_{2,3}^{(j)}\right\}=0$. Specifically, $x$ is NW restricted.

\item The set of preferences over triple $(w,x,z)$ is NSW value restricted due to $\max_{j\in N}\left\{PPM_{2,3}^{(j)}\right\}=0$, NSB value restricted due to $\max_{j\in N}\left\{PPM_{1,1}^{(j)}\right\}=1/2$ and $\max_{j\in N}\left\{PPM_{2,1}^{(j)}\right\}=1/2$, and NSM value restricted due to $\max_{j\in N}\left\{PPM_{1,2}^{(j)}\right\}=1/2$ and $\max_{j\in N}\left\{PPM_{3,2}^{(j)}\right\}=1/2$. Specifically, $z$ is NSM restricted, $w$ is both NSB restricted and NSM restricted, and $x$ is both NSB restricted and NSW restricted.
\end{itemize}

(4) With regard to triple $(x,y,z)$, the PMs corresponding to the five individuals are
$$
\begin{array}{c}
{\begin{array}{cc}
{\begin{array}{c} x\\
y\\
z
\end{array}}&
{ \left[\begin{array}{c}
\{1\}\\
\{2\}\\
\{3\}\end{array}\right],
\left[\begin{array}{c}\{1\}\\
\{3\}\\
\{2\}\end{array}\right],
\left[\begin{array}{c}\{1,2\}\\
\{3\}\\
\{1,2\}\end{array}\right],
\left[\begin{array}{c}\{2,3\}\\
\{2,3\}\\
\{1\}\end{array}\right],
\left[\begin{array}{c}\{3\}\\
\{2\}\\
\{1\}\end{array}\right].
}
\end{array}}
\end{array}
$$
Accordingly, their respective PPMs are
$$
\begin{array}{c}
\begin{array}{c}
{\begin{array}{cc}
{\begin{array}{c} x\\
y\\
z
\end{array}}&
{ \left[\begin{array}{ccc}
1 &  0 &  0 \\
0 &  1 &  0 \\
0 &  0 &  1
\end{array}\right],
\left[\begin{array}{ccc}
1 &  0 &  0 \\
0 &  0 &  1 \\
0 &  1 &  0
\end{array}\right],
\left[\begin{array}{ccc}
1/2 &  1/2 &  0 \\
0 &  0 &  1 \\
1/2 &  1/2 &  0
\end{array}\right],
\left[\begin{array}{ccc}
0 &  1/2 &  1/2 \\
0 &  1/2 &  1/2 \\
1 &  0 &  0
\end{array}\right],
\left[\begin{array}{ccc}
0 &  0 &  1 \\
0 &  1 &  0 \\
1 &  0 &  0
\end{array}\right].
}
\end{array}}
\end{array}
\end{array}
$$
\begin{itemize}
\item
The set of preferences over triple $(x,y,z)$ is best value restricted (i.e., single-caved) due to\\ $\max_{j\in N}\left\{PPM_{2,1}^{(j)}\right\}=0$. Specifically, $y$ is NB restricted.

\item The set of preferences over triple $(x,y,z)$ is NSM value restricted due to $\max_{j\in N}\left\{PPM_{1,2}^{(j)}\right\}=1/2$, and NSB value restricted due to $\max_{j\in N}\left\{PPM_{2,1}^{(j)}\right\}=0$. Specifically, $x$ is NSM restricted, and $y$ is NSB restricted.
\end{itemize}

To summarize, we obtain the following result:
\begin{itemize}
\item [(a)] All the triples in Example 1 satisfy the VR. According to Sen (1966), the individual preferences in Example 1 yields a transitive social preference under the majority rule.
\item [(b)] Both NSB and NSM are respectively satisfied by all the triples. According to Pattanaik (1970), the individual preferences in Example 1 yields a non-empty social choice set under the majority rule.
\end{itemize}

\section{Concluding remarks}

In this article, a new mathematical tool, called possibility preference map (PPM), was introduced for representing the weak orderings. Then, we used it to reformulate the value restriction and the not-strict value restriction in social choice theory. The reformulations appear elegant since they all take the form of minmax.

\vspace{0.3cm}


\end{document}